\documentclass[prl,twocolumn]{revtex4}
\usepackage{graphicx}
\begin{document}
\unitlength=1mm
\title{Experimental targeting and control  \\ of spatiotemporal chaos in nonlinear optics}
\author{L. Pastur, L. Gostiaux, U. Bortolozzo, S. Boccaletti and P.L. Ramazza}
\affiliation{Istituto Nazionale di Ottica Applicata, Largo E. Fermi 6, 50125 Firenze, Italy}
\date{\today}

\begin{abstract}
We demonstrate targeting and control over spatiotemporal chaos in
an optical feedback loop experiment. Different stationary target
patterns are stabilized in real-time by means of a two dimensional
space extended perturbation field driven by an interfaced computer
and applied in real-space to a liquid crystal display device
inserted within a control optical loop. The flexibility of the system in
switching between different target patterns is also demonstrated.

\vglue 0.3 truecm

\end{abstract}

\maketitle

Control of complex dynamics refers to a process whereby the
critical sensitivity of such dynamics to external disturbances is
capitalized in order to select a proper tiny perturbation able to
attain a desirable target behavior. In the last decade,  a series
of relevant issues such as stabilizing a given trajectory within
an infinite set of unstable periodic orbits embedded within a
chaotic attractor ({\it control of chaos}) or bringing a chaotic
trajectory to a small neighborhood of some desired locations in
phase space ({\it targeting of chaos}) have been addressed and
solved in both low and high dimensional time-chaotic dynamics
\cite{BocGreLaiManMaz00}.

\begin{figure}[tb]
\includegraphics[width=7.5cm]{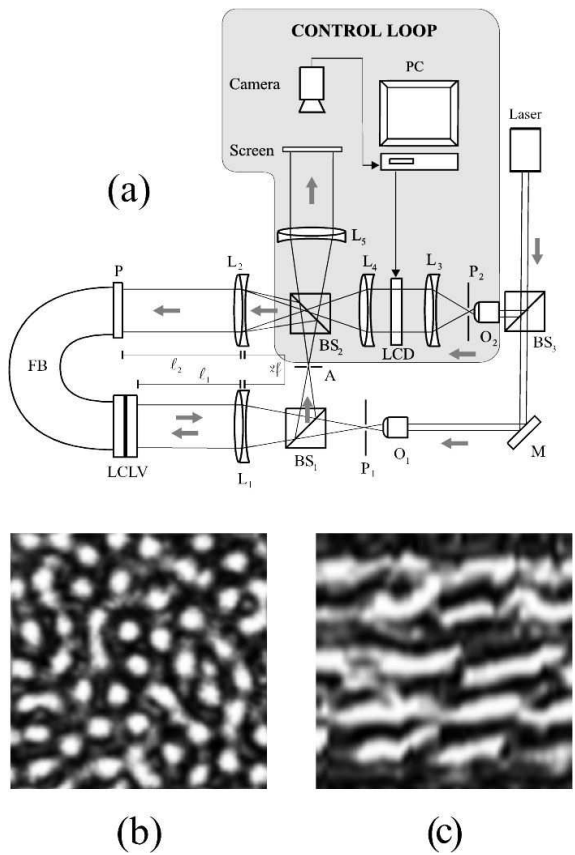}
\caption{ a) Experimental setup. Main loop: an extended laser beam
is closed through a non-linear Kerr-like medium (liquid crystal
optical valve). Instabilities develop in the transverse plane of
the beam. $M$: mirror, $O_1$: microscope objectives; P$_1$:
pinhole; $A$: aperture; BS$_1$, BS$_2$: beam splitters; LCLV:
liquid crystal light valve; L$_1$, L$_2$: lenses of focal lens
$f$; $L_5$: additional lens; FB: fiber bundle. In our experiment,
$2f-(l_1+l_2)=+90\ \rm{mm}$. Control arm: $O_2$: microscope
objectives; P$_2$: pinhole; BS$_3$: beam splitters; LCD: liquid
crystal display; L$_3$, L$_4$: lenses. The arrows indicate the
local direction of light propagation. b) Snapshot of the
uncontrolled STC state obtained for an input intensity
$I/I_{c}\sim 3.2$ ($I_c$ being the critical value for pattern
formation from the uniform state to hexagons). The pattern
intensity has been coded into a 256 levels gray scale. c) Space
(vertical)-time (horizontal) dynamical evolution of the central
vertical line of pixels of Fig. 1b.}\label{fig:setup}
\end{figure}

More recently, the interest switched to implementing control
strategies for the stabilization of space-time chaotic dynamics 
occurring in extended systems.
In this latter framework, some theoretical
attempts and numerical demonstrations have been offered for
achieving control over one and two dimensional patterns
\cite{aranson}, coupled map lattices and arrays of oscillators
\cite{parmananda}, or relevant model equations describing
universal features of systems closeby to space time bifurcations,
such as the Complex Ginzburg Landau Equation \cite{montagne} and
the complex Swift-Hohenberg Equation \cite{bleich}.

Robust and reliable experimental control over space time chaos
(STC) remains however an open problem. In the field of nonlinear optics, 
a few experimental demonstrations of pattern control and targeting have been
offered, based on Fourier filtering techniques \cite{BenKreNeuTsc00,Mamaev}. 
These methods provide
a nice efficiency for the stabilization of stationary patterns
with global symmetries, but their application to target patterns
involving complex phase relationships among different Fourier
components (such as localized patterns or selected snapshots of
STC evolving patterns) is strongly limited by the practical
difficulty of building suitable Fourier masks.

In this Letter we show the first experimental evidence of control
of STC based upon a real-space real-time feedback technique, which
is able to circumvent the above difficulties, thus allowing
stabilization and targeting of two dimensional stationary patterns
with arbitrary symmetries and shapes. This is realized by means of
perturbing fields applied directly in the real-space and in times
shorter than the characteristic time of the pattern dynamics, so
that stationary non-homogeneous patterns of arbitrary complexity in
space can be indifferently targeted and stabilized with good
efficiency. We furthermore show that our control strategy offers
dynamical flexibility in switching from one to another target
pattern, without the need of removing optical components (as e.g.
filters) in the control loop.

The experimental setup is sketched in Fig. 1a. It consists of a
main optical feedback loop (MOFL) hosting a Liquid Crystal Light
Valve (LCLV) \cite{AkhmanovNoiChaos}, and of an additional electrooptic 
control loop. The latter is essentially constituted by a videocamera, 
a personal computer driving a liquid crystal display (LCD), and a laser 
beam which traverses the LCD before being injected into the MOFL.
The LCLV operates as a Kerr-like medium, i.e., it induces on the reading 
light a phase delay proportional to the writing intensity, over a wide range 
of input intensities. This proportionality relation holds for the 
experimental parameters used in the present investigation.
The LCD display, operating in transmission, encodes linearly the gray level
images output by the PC, onto the laser beam traversing it.

When the control loop is open, a homogeneous wave is sent onto the front
face of the LCLV, and is reflected acquiring a spatial phase modulation. The
beam propagates within the MOFL from the front face of the LCLV to the input
plane of a fiber bundle, experiencing diffraction, and thus converting phase
into amplitude modulations. The fiber bundle FB just relays the intensity
distribution from the input to the output plane, which is in contact with the
LCLV rear face.

In these conditions, the dynamics within the MOFL is described by
the equation \cite{Firth,NeubeckerOppo}:
\begin{equation}
\frac{\partial \phi}{\partial t} = -\frac{1}{\tau }(\phi -\phi_0)
+ D \nabla^2 \phi + \alpha I_{fb}, \label{eq:dyn}
\end{equation}
where $\phi (x,y)$ is the phase of the optical beam at the output
of the valve, $\tau$ is a relaxation time, $D$ a diffusion
coefficient, $\alpha$  the nonlinearity strength of the LCLV,
$\phi_0$ is the working reference phase,  and $I_{fb}(x,y) $ is
the feedback intensity impinging at the rear side of the valve.

\begin{figure}[tb]
\includegraphics[width=7.5cm]{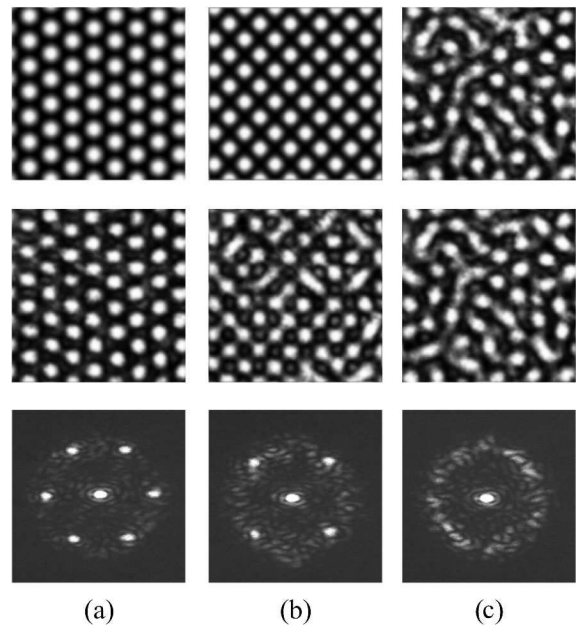}
\caption{Examples of target patterns (top row), controlled area in
the system at $\gamma =0.4$ (center row) and corresponding far
field images (bottom row) for the control trials of a perfect
hexagonal pattern (a), a square pattern (b), and a snapshot of the
uncontrolled dynamics (c).}\label{fig:snapshots}
\end{figure}

The feedback intensity, which results from the propagation of the 
beam reflected by the front side of the valve trough the MOFL, is a
nonlinear (and nonlocal) function of the phase $\phi$ \cite{Firth,NeubeckerOppo}. 
On increase of the pump intensity $I$, the homogeneous solution destabilizes,
resulting in hexagonal patterns close to threshold. If the pump is further 
increased, regular hexagons loose stability in favour of a space-time 
chaotic dynamics \cite{Firth,NeubSTC}.
Together with the pump value, another parameter of the utmost importance is the 
spatial frequency bandwidth of the system \cite{Mamaev}, controlled by the aperture 
$A$ in Fig. 1. Troughout the present paper, this bandwidth is kept fixed at 
$\simeq 1.3$ times the diffractive wavenumber of the system, which gives the 
scale of the unstable structures.

In order to achieve control over the dynamics, a fraction of the
beam traveling on the MOFL is extracted and detected by a
videocamera, which is interfaced to the PC via a frame grabber. 
The computer processes the input image, and sends a suitable
driving signal to the liquid crystal display device. The LCD
transfer function $T(x,y)$ can be written as the contribution of a
constant mean transfer coefficient $T_0$, plus a modulation signal 
$s(x,y)$. 
This modulation $s$ is chosen to be proportional to the error signal 
between the current pattern intensity $I_{fb}$ present in the system, 
and a target pattern $I_T(x,y)$:

\begin{equation}
s(x,y,t) = -\beta \left(I_{fb}(x,y,t)-I_T(x,y)\right).
\end{equation}

The digital processing operations performed by the PC include
the evaluation of the above error signal, and the calculation of 
the cross-correlation between pattern and target. The entire time series 
of pattern is also recorded on the hard disk for further off-line 
processing.

The resulting refreshing time for the above procedure is at most 200 ms, to be
compared with the characteristic time of the pattern dynamics (computed from
the decay of the autocorrelation function) which is of the order of the second. 

The LCD is illuminated with a uniform
intensity $I_0$, and the output beam is imaged onto
the rear (writing) side of the valve.
Following the above discussion, this beam will consist of a
constant term $T_0I_0$ that acts in renormalizing the valve working
point $\phi _0$  to ${\phi _0}'$, plus a modulated controlling beam 
$s I_0$. 

The equation of motion when the control loop is closed is therefore:

\begin{equation}
\frac{\partial \phi}{\partial t} = -\frac{1}{\tau } (\phi -\phi
_0^\prime) + D\Delta \phi + \alpha \left(I_{fb}-\gamma
(I_{fb}-I_T)\right). \label{eq:control}
\end{equation}

\begin{figure}[tblr]
\includegraphics[width=5.5cm]{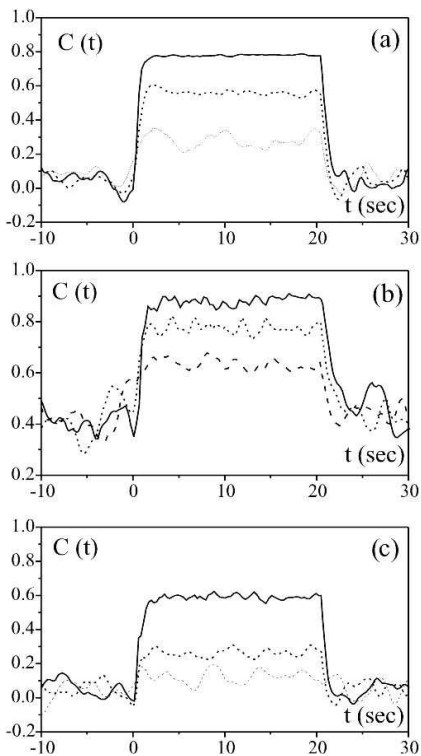}
\caption{Correlation function $C(t)$ (a,c,e) and amount of power
$p(t)$ injected within the control arm (b,d,f) {\it vs.} time (in
sec.). The two quantities are defined in the text. In all cases
the continuous line refers to $\gamma=0.4$, the dotted line to
$\gamma=0.2$, and the dashed line to $\gamma=0.1$. (a,b) the
target is a perfect hexagonal pattern; (c,d) the target is a
snapshot of the uncontrolled dynamics; (e,f) the target is a
perfect square pattern.  $I/I_c \simeq 3.2$.}\label{fig:gamma}
\end{figure}

where $\gamma \equiv \beta I_0$.

We initially set the light intensity at the input of the feedback
loop to be $I/I_{c}\sim 3.2$, where $I_c$ is the critical value
for pattern formation from the uniform state to hexagons.
In these conditions, the
uncontrolled evolution brings the system to display a time
evolving, spatially disordered pattern, where many 
defects are continuously created and annihilated within an hexagonal-like
pattern, thus generating STC \cite{NeubSTC}. A typical snapshot of the
uncontrolled dynamics is reported in Fig. 1b. Fig. 1c reports the space-time
dynamical evolution of the central vertical line of pixels in Fig. 1b, showing
how the uncontrolled dynamics evolves within STC, with a non stationary
complex local dynamics and a decaying spatial correlation.

\begin{figure}[tb]
\includegraphics[width=6.5cm]{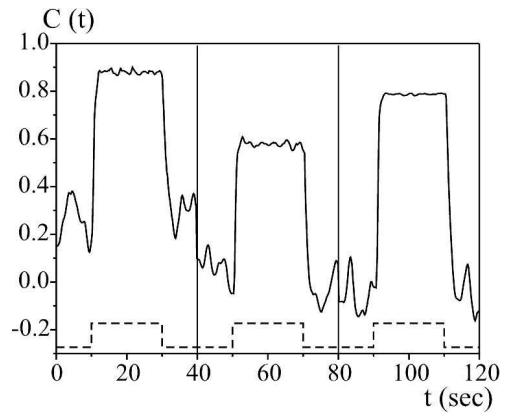}
\caption{Correlation function $C(t)$ (see text for definition)
{\it vs.} time during the sequential control trial at $\gamma
=0.4$. The target patterns are a snapshot of the uncontrolled
dynamics ($10 \leq t < 30$ sec.), a square pattern ($50 \leq t <
70$ sec.) and a hexagonal pattern ($90 \leq t < 110$ sec.). The
dashed line indicates the switching on/off times. The vertical
lines separates the three domains in time in which a different
pattern is taken as target for the control. In each time domain,
the correlation is calculated using the corresponding target
pattern.}\label{fig:dyn}
\end{figure}

Starting from these conditions, three different target patterns
are selected, namely perfect hexagons, squares and a particular
snapshot of the uncontrolled dynamical evolution (shown in the top
row of Fig. \ref{fig:snapshots}). 

Perfect stationary hexagons are a stable solution close to the pattern 
formation threshold; they are destabilized when the pump is increased, 
resulting in the space-time chaotic dynamics here considered. Therefore, 
use of hexagons as a target tests the ability of the method to control 
an unstable solution.

On the other hand,
using as target a snapshot of the uncontrolled dynamics assesses
the robustness of the method to freeze a given natural state
of the uncontrolled dynamics, in the very same spirit as the so
called targeting of chaos \cite{BocGreLaiManMaz00}. Finally,
squares  are never spontaneously selected by the system without
control, and therefore they serve us to assess the ability of the
control strategy to force the appearance of an arbitrary symmetry.
The results obtained for $\gamma =0.4$ in the three cases are
reported in the center row of Fig.\ref{fig:snapshots}, indicating
that the control procedure is successful in all cases. The
patterns are conveniently stabilized over a selected area of
$128\times 128$ pixels (about 10 pattern wavelengths) in the system.
The bottom
row of Fig.\ref{fig:snapshots} shows the far field images of the
controlled dynamics. While hexagonal and square patterns have a
rather simple global symmetry (thus allowing for an easy
implementation of Fourier filtering techniques, like the one of
Ref. \cite{BenKreNeuTsc00}), the target STC snapshot involves 
the presence of a complicate power spectrum. The fabrication 
of Fourier masks reproducing the amplitude and phase of these patterns 
appears extremely difficult in experiments.

Our control method circumvents such practical
difficulties, and effectively stabilizes such complex Fourier
patterns. Hence, it configures as the most reliable choice for the
stabilization and targeting of two dimensional stationary structures
with arbitrary symmetries and shapes.

To evaluate quantitatively the control ability of our method, we use 
the time dependent correlation function
$C(t)=<I_{fb}(\textbf{r},t)\cdot I_T(\textbf{r})>_{\textbf{r}}$
between the instantaneous pattern and the target one 
($<...>_{\textbf{r}}$ denotes a spatial average).
We also measure the amount of power $p(t)=\gamma
[<(I_{fb}(\textbf{r},t)-I_T(\textbf{r}))^2>_{\textbf{r}}]^{1/2}$
injected within the control arm for controlling the target
pattern.

The correlations $vs$ time are shown in Fig.\ref{fig:gamma} for
increasing values of $\gamma $ for the control task of a perfect
hexagonal pattern (a), a snapshot of the uncontrolled dynamics
(b), and a square pattern (c).  At lower values of $\gamma$
($\gamma=0.1,0.2$), there is already a partial control of the dynamics, 
though several deviations of the patterns around the target ones 
still remain. This is reflected by the rather large fluctuations of 
$C(t)$ visible in Fig. 3. The correlation value increases with $\gamma $
up to $\gamma =0.4$, when a high degree of control is achieved in all cases.   
As $\gamma$ increases, the transient time before reaching control over the
target pattern decreases. On the opposite, when control is switched off, the
relaxing time is determined by the characteristic time of the STC
decorrelation, and is therefore independent of $\gamma$. 
The value of the control power $p(t)$ needed to achieve control is in 
the range $0.1 - 0.2$ for the data reported.  

Notice that the maximum
correlation is attained for the case of the STC snapshot. This is 
due to the fact that in this case the target pattern is a specific
configuration of the natural evolution of the dynamics. In the cases of 
squares and hexagons, instead, the target patterns 
are are digitally generated trying to reproduce at best the natural profile of
the spots present in the system; but this procedure is intrinsically imperfect,
because the exact profile of the spots is unknown. 
The mismatch between the
profiles of the targets and the patterns lowers the correlation levels, with
respect to the values that would be obtainable in the ideal case, in which the
profile of the solution is perfectly reproduced by the target image.

It is worth observing that the correlation when control is off in Fig. 3c 
does not decay to zero; this is due to the fact that our uncontrolled STC 
dynamics gives rise to a non zero mean field. This point can be qualitatively 
appreciated from inspection of Fig. 1c: the pattern has a certain degree of 
"phase rigidity", i.e., even if there are chaotic fluctuations, bright (dark) 
areas remain more or less bright (dark) for most of time.
Similar properties have been observed experimentally and discussed in various
other cases of space extended systems giving rise to STC dynamics \cite{vari}.

Finally, we demonstrate that our control technique is greatly
flexible in dynamically switching between different target
patterns. For this purpose, we prepare a time-dependent target
pattern formed by the ordered sequence of a snapshot of the
uncontrolled dynamics, a square pattern and a hexagonal pattern,
each one presented to the system for a time $T=20$ sec., and
intermingled with periods of  $T=20$ sec. of time in which the
system is left uncontrolled to reset the original state of STC.
The results are shown in Fig. \ref{fig:dyn}, where one sees that
the system is able to attain each one of the target pattern in the
sequence for the same value of $\gamma=0.4$, as well as to switch
between the different patterns. Notice that each target pattern in
the sequence produces a different state in the correlation,
depending upon its specific instability features within the
uncontrolled STC regime. The maximal correlation is again obtained
for the snapshot of the uncontrolled dynamics, since this pattern
represents a specific state compatible with the uncontrolled
dynamics.

In conclusion we have demonstrated targeting and control over
spatiotemporal chaos in an optical feedback loop experiment.
Stationary target patterns with arbitrary symmetry and complexity
are stabilized in real-time by means of a two dimensional space
extended perturbation field at low cost in terms of the
perturbation amplitude. The flexibility of the system in switching
between different target patterns has also been demonstrated.

Work partly supported by EU Contract HPRN-CT-2000-00158, and
MIUR-FIRB project n. RBNE01CW3M-001.

\end{document}